\documentclass{JHEP}

\title{Borel summation and momentum-plane analyticity in 
perturbative QCD}

\author{Irinel Caprini\\
National Institute of Physics and Nuclear Engineering\\
Bucharest POB MG-6, R-76900 Romania\\
E-mail: \email{caprini@theor1.ifa.ro}}

\author{Matthias Neubert\\
Stanford Linear Accelerator Center, Stanford University\\
Stanford, California 94309, U.S.A.\\
E-mail: \email{neubert@slac.stanford.edu}}

\abstract{We derive a compact expression for the Borel sum of a QCD
amplitude in terms of the inverse Mellin transform of the
corresponding Borel function. The result allows us to investigate the
momentum-plane analyticity properties of the Borel-summed Green
functions in perturbative QCD. An interesting connection between the
asymptotic behaviour of the Borel transform and the Landau
singularities in the momentum plane is established. We consider for
illustration the polarization function of massless quarks and the
resummation of one-loop renormalon chains in the large-$\beta_0$
limit, but our conclusions have a more general validity.}


\preprint{SLAC-PUB-8059\\
hep-ph/9902244}

\begin{document}

\section{Introduction}

It is well-known that the momentum-plane analyticity properties
imposed by causality and unitarity are not automatically satisfied by
the  correlation functions calculated in perturbative QCD. This is
actually a more general feature of approximate solutions of physical
equations. For instance, unexpected analytic properties  were
discovered for the perturbative solutions in potential theory
\cite{KhWu}. In QCD, the problem is more severe due to
confinement. Indeed, as shown in \cite{Oehme}, in a confined theory
the poles and branch points of the true Green functions are generated
by the physical hadron states in the unitarity relation, and no
singularities  related to the underlying quark and gluon degrees of
freedom should appear. On the other hand, in perturbation theory the
branch points  of the amplitudes correspond to the production of free
quarks and gluons.  Other, more complicated singularities make their
appearance if one goes beyond naive perturbation theory.  For
instance, renormalization-group improved expansions have unphysical
space-like singularities (Landau poles or cuts), while the 
introduction of non-perturbative contributions such as vacuum
condensates in the Operator Product Expansion (OPE) produces poles
and branch cuts at the origin \cite{SVZ}. Both types of singularities
are absent from the exact QCD amplitudes.

Recently, the contribution of large orders of perturbation theory
related to the so-called renormalons \cite{tHof} have  been
investigated by several authors, using in particular the method of
Borel summation \cite{Muel}--\cite{BBK} (see \cite{Martin} for a 
review and further references). In general, these resummations 
introduce additional momentum-plane singularities in QCD Green 
functions, reflecting the presence of the ultraviolet (UV) and 
infrared (IR) renormalons in their Borel transforms. But despite 
of the large number of works devoted to Borel summation in 
perturbative QCD, the momentum-plane analyticity of the Borel sum 
did not receive particular attention so far. The reason presumably 
lies in the fact that perturbation theory by itself is not complete 
and must be supplemented by non-perturbative terms to build up the 
exact theory. It is therefore not surprising that theoretical
predictions based on perturbation theory are plagued by bad
analyticity properties absent from the true physical amplitudes. 
Nevertheless, given that analyticity is the key property allowing 
the continuation of theoretical predictions from Euclidian to 
Minkowskian kinematics relevant to hadronic observables, we believe
an investigation of the momentum-plane analyticity of the Borel sum 
in perturbative QCD is of both theoretical and practical interest.
  
The importance of the problem was emphasized some time ago by Khuri
\cite{Khuri}, who noted the mathematical possibility of a Borel sum
with  analytic and asymptotic properties in the momentum plane
different from  those of the individual terms in the perturbative
expansion. (Khuri also  quotes unpublished results by J.P.~Eckmann and
T.~Spencer illustrating  this possibility.)  The deep connection
between momentum-plane analyticity and Borel summability was revealed
by 't~Hooft \cite{tHof}, who showed that the presence of resonances
and multi-particle thresholds on the time-like axis is in conflict
with the mathematical  conditions required for the existence of the
Borel sum. More recently, the relation between the asymptotic
behaviour of the Borel transform and the momentum-plane singularities
of the QCD amplitudes was discussed in \cite{BBB}.
Renormalization-group invariance and the Landau pole in the running
coupling $\alpha_s(Q^2)$ are the essential ingredients of this
analysis. Attempts  to remove this unphysical singularity in the
context of dispersion theory with an IR-regular running coupling were 
made in \cite{DoMa}. This procedure raised much interest, since it 
generates power corrections of a type not present in the OPE 
\cite{Grun1,AkZa} (see, however, the discussion in \cite{BBK2}). 
Further clarifications of the connection between the Landau 
singularity of the running coupling and the IR renormalons of the 
Borel transform are desirable in this context.

In the present paper, we discuss some aspects of the momentum-plane
analyticity of Borel-summed Green functions in  perturbative QCD.
To this end, we consider the polarization function for massless quarks
and calculate its Borel sum  at complex values of the momentum 
transfer. The result we derive, written in a compact form in terms of a
``distribution function'' (i.e., an inverse Mellin transform of the
Borel function),  allows us to investigate the singularities generated
by the process of Borel summation. Since we consider the massless
case,  branch points at $s=0$ will be present in the time-like region,
which  of course do not reproduce the true hadronic thresholds. Our 
main concern, however, is the appearance of unphysical (Landau)
singularities  in the space-like region.  Our discussion is kept at a
rather formal level, although  reference to the specific case of a
single renormalon chain (corresponding to the so-called 
large-$\beta_0$ limit \cite{Matt1,Martin})
is made to illustrate various results and assumptions. In
Section~\ref{sec:2}, we start with a brief review of the integral
representations of the Borel function in terms of its inverse Mellin
transforms. Two different types of inverse  Mellin transforms (called
distribution functions) were introduced in \cite{Matt1} and
\cite{BBB}. We indicate the mathematical conditions under which these
functions exist, the relations among them, and their connection with 
renormalons. An important finding is that the inverse
Mellin transform of the Borel function associated with the 
polarization amplitude is, in general, a piece-wise analytic function, 
with two pieces defined inside and outside a circle in the complex 
plane, each of them being analytic except for a cut on the real axis. 
In Section~\ref{sec:3}, we derive a compact expression for the 
polarization function at complex values of the momentum variable in 
terms of the distribution function defined in \cite{Matt1}. The same 
technique is applied for obtaining an expression for the Borel 
sum of Minkowskian quantities, using as input the standard result 
for the Borel transform of these quantities obtained by analytic 
continuation. In Section~\ref{sec:4}, we discuss the analyticity 
properties of the polarization amplitude in the momentum plane. These 
properties depend on the prescription used for treating the IR 
renormalon singularities of the Borel transform. We adopt a 
principal-value prescription, which leads to amplitudes without 
space-like singularities outside the ``Landau region'' 
(the region between 0 and the location of Landau singularities on 
the space-like $s$ axis in the renormalization-group improved 
perturbative expansions of correlators in QCD) and is therefore 
consistent with the analyticity requirements imposed by causality 
and unitarity. The analytic continuation to low momenta reveals 
explicitly the connection between the Landau singularities and the 
piece-wise character of the inverse Mellin transform, which in turn 
is determined by the asymptotic behaviour of the Borel function. Our 
conclusions are given in Section~\ref{sec:5}. The paper has an 
Appendix, in which we present a method for deriving relations 
between the various distribution functions based on Parseval's 
theorem for Mellin transforms \cite{Tit}.

\section{Borel transforms and distribution functions}
\label{sec:2}

Consider the current--current correlation function $\Pi(q^2)$ defined 
by
\begin{equation}\label{2point}
   -4\pi^2 i\int{\rm d}^4 x\,e^{i q\cdot x}\,\langle 0|\,T\,
   \{ V^\mu(x),V^\nu(0) \}\,|0\rangle
   = (q^\mu q^\nu - g^{\mu\nu} q^2)\,\Pi(q^2) \,,  
\end{equation}
where $V^\mu=\bar q\gamma^\mu q$ is the conserved vector current 
for massless quarks. We define the corresponding Adler function
\begin{equation}\label{Ddef}
   D(s) = -s\,\frac{{\rm d}\Pi(s)}{{\rm d}s} \,,\quad s=q^2 \,,
\end{equation}
which is UV finite. The function $\Pi(s)$ can be obtained from $D(s)$ 
by logarithmic integration, yielding
\begin{equation}\label{Pidef}
   \Pi(s) = -\int\limits^s\!{\rm d}\ln(-s')\,D(s')
   + \mbox{const.} \,,
\end{equation}
where the integration is along a contour in the complex $s'$ plane 
ending at the point $s$ and not encountering the singularities of 
the integrand. General causality and unitarity properties of QCD Green 
functions imply that $\Pi(s)$ and $D(s)$ are real analytic functions 
in the complex $s$ plane (i.e., $\Pi(s^*)=\Pi^*(s)$ and 
$D(s^*)=D^*(s)$), cut along the positive real axis from the threshold 
for hadron production at $s=4 m_\pi^2$ to infinity. The analyticity 
properties of the exact Adler function $D(s)$ are summarized in the 
dispersion relation
\begin{equation}\label{dispreld}
   D(s) = \frac{s}{\pi} \int\limits_{4m_\pi^2}^\infty\!{\rm d}s'\,
   \frac{\mbox{Im}\,\Pi(s'+i\epsilon)}{(s'-s)^2} \,,
\end{equation}
where the spectral function ${\rm Im}\,\Pi(s+i\epsilon)$ is 
non-negative. The function $\Pi(s)$ satisfies a once-subtracted 
dispersion relation similar to (\ref{dispreld}).

Writing the one-loop running coupling constant as
\begin{equation}\label{asrng}
   \alpha_s(-s) = \frac{4\pi}{\beta_0\ln(-s/\Lambda^2)} \,,
\end{equation}
where $\beta_0=\frac{11}{3} N_c - \frac 23 n_f$ is the first coefficient 
of the $\beta$ function and $\Lambda$ the QCD scale parameter, one finds 
that the first terms in the renormalization-group improved perturbative 
expansions of the functions $D(s)$ and $\Pi(s)$ are given by
\begin{eqnarray}\label{Drengr}
   D(s) &=& 1 + \frac{4}{\beta_0\ln(-s/\Lambda^2)} + \dots \,,
    \nonumber\\
   \Pi(s) &=& k - \ln(-s/\Lambda^2)
    - \frac{4\ln\ln(-s/\Lambda^2)}{\beta_0} + \dots \,,
\end{eqnarray}
where $k$ is a constant.  These
expressions are  derived using the renormalization-group equations in
the perturbative region $|s|\gg\Lambda^2$. Their analytic continuation
to low energy contains unphysical singularities on the  space-like
axis, which are absent from the exact amplitudes. Specifically,
$D(s)$ has a Landau pole at $s=-\Lambda^2$, and $\Pi(s)$  has a Landau
cut along the interval  $-\Lambda^2<s<0$.  In \cite{DoMa} it was
proposed to restore the correct analyticity  properties by a
redefinition of the coupling (\ref{asrng}) using   a dispersion
relation. In the present paper, we take a different
perspective and analyse  the impact of large orders of perturbation
theory on the analyticity properties of the correlation functions.

The structure of high-order contributions to the perturbative expansion 
of the Adler function was investigated by many authors. The expansion 
coefficients are known to exhibit a factorial growth, indicating that the 
perturbation series has zero radius of convergence \cite{Bene,Broa}. If 
one formally applies the Borel method to sum the series, this growth is 
reflected in singularities of the Borel transform on the real axis. Let 
us write the formal Laplace integral expressing $D(s)$ in terms of its 
Borel transform in the large-$\beta_0$ limit as \cite{Martin}
\begin{eqnarray}\label{Laplace}
   D(s) &=& 1 + \frac{1}{\beta_0} \int\limits_0^\infty\!{\rm d}u
    \left( \frac{-s e^{-C}}{\mu^2} \right)^{-u}
    \exp\left( -\frac{4\pi u}{\beta_0\alpha_s(\mu^2)} \right)
    \widehat B_D(u) \nonumber\\
   &=& 1 + \frac{1}{\beta_0} \int\limits_0^\infty\!{\rm d}u
    \left( \frac{-s}{\Lambda_V^2} \right)^{-u} \widehat B_D(u) \,,
\end{eqnarray}
where $C$ is a scheme-dependent constant, which takes the value 
$C=-5/3$ in the $\overline{\rm MS}$ scheme and $C=0$ in the V scheme, 
and $\Lambda_V^2=e^{-C}\Lambda^2$ is a scheme-independent combination 
of this constant with the QCD scale parameter. The corresponding 
representation for $\Pi(s)$ is 
\begin{equation}\label{Laplace1}
   \Pi(s) = k -\ln\left( \frac{-s}{\Lambda_V^2} \right)
   + \frac{1}{\beta_0} \int\limits_0^\infty\!{\rm d}u \left( 
   \frac{-s}{\Lambda_V^2} \right)^{-u} \widehat B_\Pi(u) \,.
\end{equation}
We have used the renormalization-group invariance to isolate the 
dependence on the momentum variable $s$, so that the Borel 
transforms $\widehat B_D(u)$ and $\widehat B_\Pi(u)$ are scale and 
scheme independent and also do not depend on $s$. Note that the 
definition (\ref{Ddef}) implies the relation 
\begin{equation}\label{BDBPi}
   \widehat B_\Pi(u) = \frac{1}{u}\,\widehat B_D(u) \,.
\end{equation}

The Borel transforms $\widehat B_D(u)$ and $\widehat B_\Pi(u)$ have 
singularities on the real axis in the complex $u$ plane. Some of them, 
the so-called IR renormalons, are situated along the integration
contour in (\ref{Laplace}) and (\ref{Laplace1}).\footnote{The function 
$\widehat B_\Pi(u)$ has, in addition, a pole at $u=0$ related to the 
short-distance behaviour of the current correlator. This pole is 
removed by standard UV renormalization.}
The precise nature of these singularities is only known for some 
specific cases. For instance, the first IR and UV singularities are 
branch points located at $u=2$ and  $u=-1$, respectively, with a 
universal nature determined by renormalization-group coefficients 
\cite{Muel,BBK}. Information on the other singularities is 
only available from calculations performed in the large-$\beta_0$ 
limit, which predict the correct locations of the singularities but 
not their nature (in general, the large-$\beta_0$ approximation yields 
pole singularities rather than branch points). Specifically, the 
expression for the Borel transform of the Adler function is 
\cite{Bene,Broa}
\begin{equation}\label{BDu}
   \widehat B_D(u) = \frac{128}{3(2-u)}\,\sum_{k=2}^\infty\,
   \frac{(-1)^k\,k}{\big[ k^2-(1-u)^2 \big]^2} \,.
\end{equation}
The presence of IR renormalons renders the perturbation 
series not Borel summable and the Laplace integrals ill defined. In 
order to evaluate them, a prescription for avoiding the 
singularities must be specified. A question we address in this paper is 
whether additional information can be employed to prefer one choice of 
a prescription over another. We shall argue that momentum-plane 
analyticity might be a helpful guiding principle in this respect.

We also consider Minkowskian quantities such as the finite-energy 
moments of the spectral function of $\Pi(s)$ defined as
\begin{equation}\label{Mkdef}
   M_k = \frac{k+1}{\pi\,s_0^{k+1}} 
   \int\limits_{4m_\pi^2}^{s_0}\!{\rm d}s\,s^k\,
   \mbox{Im}\,\Pi(s+i\epsilon) 
   = -\frac{k+1}{2\pi i\,s_0^{k+1}}
   \!\oint\limits_{|s|=s_0}\!\!{\rm d}s\,s^k\,\Pi(s) \,,
\end{equation}
where $s_0>4 m_\pi^2$ is an arbitrary cutoff, and we have applied the 
Cauchy relation for the true, physical polarization function $\Pi(s)$ 
to pass from the integral along the cut to a contour integral along a 
circle in the complex $s$ plane. Combinations of such moments enter 
the prediction for physical observables such as the hadronic decay 
rate of the $\tau$-lepton \cite{Matt3}, and some tests of QCD related 
to the dependence under variation of $s_0$ are discussed in 
\cite{Fpich,MG}. The Borel representation for Minkowskian quantities 
such as the spectral moments $M_k$ can be derived starting from 
the contour integral written in the second relation in (\ref{Mkdef}) 
and inserting the expression (\ref{Laplace1}) for $\Pi(s)$. The 
result is
\begin{equation}\label{momb}
   M_k = 1 + \frac{1}{\beta_0}\,
   \int\limits_0^\infty\!\mbox{d}u \left( \frac{s_0}{\Lambda_V^2}
   \right)^{-u} \widehat B_{M_k}(u) \,,\quad
   s_0 > \Lambda_V^2 \,,
\end{equation}
where \cite{Matt3}
\begin{equation}\label{BdBPi3}
   \widehat B_{M_k}(u) = \frac{k+1}{k+1-u}\,
   \frac{\sin\pi u}{\pi u}\,\widehat B_D(u) \,.
\end{equation}

Next we define the ``distribution functions'' coresponding to the 
above Borel transforms. We consider first the Borel transform 
$\widehat B_D(u)$ of the Adler function and assume that it is 
analytic in the strip $u_1<\mbox{Re}\,u<u_2$, where $u_1=-1$ and 
$u_2=2$ are the positions of the first UV and IR renormalons, 
respectively. Let us further assume that the following $L^2$ condition 
holds:
\begin{equation}\label{L2B}
   \frac{1}{2\pi i} \int\limits_{u_0-i\infty}^{u_0+i\infty}\!
   {\rm d}u\,|\widehat B_D(u)|^2 < \infty \,,
\end{equation}
where $u_0\in[u_1,u_2]$. Then the inverse Mellin transform of 
$\widehat B_D(u)$ can be defined as \cite{Tit} 
\begin{equation}\label{wDn}
   \widehat w_D(\tau) = \frac{1}{2\pi i}
   \int\limits_{u_0-i\infty}^{u_0+i\infty}\!{\rm d}u\,
   \widehat B_D(u)\,\tau^{u-1} \,.
\end{equation}
The condition (\ref{L2B}) is satisfied by the Borel transform 
calculated in the large-$\beta_0$ limit, given in (\ref{BDu}). 
In this case, the function $\widehat w_D(\tau)$ was first introduced 
in \cite{Matt1}, where its physical interpretation as the distribution 
of the internal gluon virtualities in Feynman diagrams was pointed 
out. Relation (\ref{wDn}) can be inverted to give 
\begin{equation}\label{wninv}
   \widehat B_D(u) = \int\limits_0^\infty\!{\rm d}\tau\,
   \widehat w_D(\tau)\,\tau^{-u} \,,
\end{equation}
and the following completeness condition holds \cite{Tit}:
\begin{equation}\label{compl}
   \frac{1}{2\pi} \int\limits_{u_0-i\infty}^{u_0+i\infty}\!{\rm d}u\,
   |\widehat B_D(u)|^2 =\int\limits_{0}^\infty
   \frac{{\rm d}\tau}{\tau}\,|\tau^{1-u_0}\,\widehat w_D(\tau)|^2 \,.
\end{equation}
The integral relation (\ref{wninv}) defines the function 
$\widehat B_D(u)$ in a strip parallel to the imaginary axis with 
$u_1<\mbox{Re}\,u<u_2$.

The distribution function $\widehat w_D(\tau)$ can be calculated from 
(\ref{wDn}) by closing the integration contour along a semi-circle at
infinity in the $u$ plane and applying the theorem of residues for the 
singularities of $\widehat B_D(u)$ located inside the integration 
domain. For $\tau<1$ the contribution from the 
semi-circle at infinity vanishes if the contour is closed in the 
right half of the $u$ plane, while for $\tau>1$ the contour must be
closed in the left half plane. One thus obtains different 
expressions for the distribution function depending on whether
$\tau<1$ or $\tau>1$, which we shall denote by $\widehat 
w_D^{(<)}(\tau)$ and $\widehat w_D^{(>)}(\tau)$, respectively. 
By Cauchy's theorem, these functions are given by the contribution of 
the residues of the IR and UV renormalons, respectively. For instance,
$\widehat w_D^{(<)}(\tau)$ has the expression
\begin{equation}\label{wn1}
   \widehat w_D^{(<)}(\tau) = \frac{1}{2\pi i}\int_{\cal C_+}\!
   {\rm d}u\,\widehat B_D(u)\,\tau^{u-1} - \frac{1}{2\pi i}
   \int_{\cal C_-}\!{\rm d}u\,\widehat B_D(u)\,\tau^{u-1} \,,
\end{equation}
where the contours ${\cal C_+}$ and ${\cal C_-}$ are lines slightly 
above and below the positive real axis, respectively. A similar 
expression applies for the function $\widehat w_D^{(>)}(\tau)$, where 
the integration lines are now parallel to the negative real axis. In 
the large-$\beta_0$ limit, these functions are given by \cite{Matt1} 
\begin{eqnarray}\label{wDfun}
   \widehat w_D^{(<)}(\tau) &=& \frac{32}{3} \left\{ \tau\left(
    \frac74 - \ln\tau \right) + (1+\tau)\Big[ L_2(-\tau) + \ln\tau
    \ln(1+\tau) \Big] \right\} \,, \nonumber\\
   \widehat w_D^{(>)}(\tau) &=& \frac{32}{3} \left\{ 1 + \ln\tau
    + \left( \frac34 + \frac12 \ln\tau \right) \frac{1}{\tau}
    + (1+\tau)\Big[ L_2(-\tau^{-1}) - \ln\tau \ln(1+\tau^{-1}) \Big]
    \right\} \,, \nonumber\\
\end{eqnarray}
where $L_2(x)=-\int_0^x {{\rm d}t\over t}\ln(1-t)$ is the dilogarithm.
The above expressions are analytic in the complex $\tau$ plane, with
no singularities other than branch cuts along the negative real
axis. Together they define a function $\widehat w_D(\tau)$ that
is piece-wise analytic in the cut $\tau$ plane, with different 
functional expressions for $|\tau|<1$ and $|\tau|>1$. 

The fact that the function $\widehat w_D(\tau)$ is piece-wise 
analytic will have interesting consequences in our analysis below. 
One might ask whether these analytic properties are valid in 
general, i.e., beyond the large-$\beta_0$ limit. It is important in
this context that we consider the analytic continuations of the function  
$\widehat w_D^{(<)}(\tau)$ defined in (\ref{wn1}) and of the function  
$\widehat w_D^{(>)}(\tau)$ defined in a similar way, and not
the original definition (\ref{wDn}). The equivalence between the two 
results is valid only for real $\tau$. By writing $\tau^u
=\exp(u\ln\tau)$, and noticing that the residues of the renormalon 
singularities are real, we obtain from (\ref{wn1}) real values for
positive $\tau$. However, a branch point at $\tau=0$ and a cut at 
negative $\tau$ may appear due to the logarithm. In general, the 
singularities of the Borel transform are expected to be branch 
points (rather than poles) located on the real $u$ axis. The 
inverse Mellin transforms 
of functions with branch points have a similar structure as those 
of functions with pole singularities \cite{Bate}: they are
piece-wise analytic functions, composed of two pieces analytic in the
domains $|\tau|<1$ and $|\tau|>1$, with a cut along the
negative real axis. Therefore, this seems to be the most general 
analytic structure for the function $\widehat w_D(\tau)$, assuming that
the asymptotic behaviour of the Borel transform is in accord with the 
norm condition (\ref{L2B}). 

The application of the above procedure for the polarization function
itself requires some care due to the additional pole at $u=0$ of the 
Borel transform $\widehat B_\Pi(u)$. An expression consistent with the 
definitions (\ref{Ddef}) and (\ref{Pidef}) is obtained if we define 
the inverse Mellin transform $\widehat w_\Pi(\tau)$ in terms of an 
integral along the line $\mbox{Re}\,u=u_0$ with $u_0>0$. Then, as 
proven in the Appendix, the result is
\begin{equation}\label{wdwpi}
   \widehat w_\Pi(\tau) = \frac{1}{\tau} \int\limits_0^\tau\!
   {\rm d}x\,\widehat w_D(x) \,.
\end{equation}
In the large-$\beta_0$ limit, the explicit expression for this 
function is \cite{Matt1}
\begin{equation}
   \widehat w_\Pi(\tau) = \frac{16}{3} \left\{ 1 - \ln x
   + \frac{x}{2}\,(5-3\ln x) + \frac{(1+x)^2}{x}
   \Big[ L_2(-x) + \ln x\ln(1+x) \Big] \right\} \,,
\end{equation}
where $x=\tau$ if $\tau<1$, and $x=1/\tau$ if $\tau>1$. It follows
that $\widehat w_\Pi^{(>)}(\tau)=\widehat w_\Pi^{(<)}(1/\tau)$. 

Consider now the case of Minkowskian quantities such as the spectral 
moments $M_k$. In this case, the presence of the factor 
$\sin\pi u$ in the Borel transforms in (\ref{BdBPi3}) affects their 
asymptotic behaviour for large $u$ and invalidates the $L^2$ condition 
(\ref{L2B}). Therefore, in general an inverse Mellin transform 
defined as in (\ref{wDn}) does not exist for Minkowskian quantities.
This observation suggests to extract the sine factor from the Borel
transform and consider the inverse Mellin transform of the remaining 
expression. Actually, such a definition can be considered not only 
for Minkowskian quantities but for any Borel transform 
$\widehat B_X(u)$ \cite{BB}. We define a function
\begin{equation}\label{Bb}
   \widehat b_X(u) = \frac{\widehat B_X(u)}{\sin\pi u}
\end{equation}
and assume that it obeys the $L^2$ condition
\begin{equation}\label{L2B1}
   \frac{1}{2\pi i} \int\limits_{u_0-i\infty}^{u_0+i\infty}\!
   {\rm d}u\,|\widehat b_X(u)|^2 = \frac{1}{2\pi i}
   \int\limits_{u_0-i\infty}^{u_0+i\infty}\!{\rm d}u\,
   \left| \frac{\widehat B_X(u)}{\sin\pi u} \right|^2 < \infty \,,
\end{equation}
where $u_0\ne 0$. In the case of Minkowskian quantities, the sine 
function in the denominator of (\ref{L2B1}) compensates the factor of 
$\sin\pi u$ appearing in the Borel transform. (In the case of 
Euclidean quantities the sine factor brings an additional improvement 
in the convergence of the integrals.) Provided the condition 
(\ref{L2B1}) is satisfied, we can then define the inverse Mellin 
transform $\widehat W_X(\tau)$ as
\begin{equation}\label{wbb}
   \widehat W_X(\tau) = \frac{1}{2\pi i}
   \int\limits_{u_0-i\infty}^{u_0+i\infty}\!{\rm d}u\,
   \widehat b_X(u)\,\tau^{u-1} = \frac{1}{2\pi i}
   \int\limits_{u_0-i\infty}^{u_0+i\infty}\!{\rm d}u\,
   \frac{\widehat B_X(u)}{\sin\pi u}\,\tau^{u-1} \,.
\end{equation}
This definition was introduced in a more physical context in 
\cite{BBB}. Relation (\ref{wbb}) can be inverted to give
\begin{equation}\label{wbbinv}
   \widehat b_X(u) = \int\limits_0^\infty\!{\rm d}\tau\,
   \widehat W_X(\tau)\,\tau^{-u} \,,
\end{equation}
which together with (\ref{Bb}) defines the Borel function 
$\widehat B_X(u)$ along a line $\mbox{Re}\,u=u_0>0$, where $u_0$ can be 
taken arbitrarily small. Of course, if the Borel transform 
$\widehat B_X(u)$ itself satisfies an $L^2$ condition like in 
(\ref{L2B}), then also the normal inverse Mellin transform 
$\widehat w_X(\tau)$ defined as in (\ref{wDn}) exists. In this case, 
one can prove the relation \cite{Matt2,BBB}
\begin{equation}\label{rel}
   \widehat W_X(\tau) = \frac{1}{\pi} \int\limits_0^\infty\!
   {\rm d}x\,\frac{\widehat w_X(x)}{x+\tau} \,,
\end{equation}
which we derive in the Appendix using Parseval's theorem. The 
representation (\ref{rel}) shows that, unlike the distribution 
functions $\widehat w_D(\tau)$ and $\widehat w_\Pi(\tau)$, which are 
piece-wise analytic in the $\tau$ plane, the modified inverse Mellin 
transforms $\widehat W_D(\tau)$ and $\widehat W_\Pi(\tau)$ are real 
analytic functions in the entire complex $\tau$ plane cut along the 
real negative axis. (The explicit expression for the function 
$\widehat W_D(\tau)$ can be found in the Appendix of \cite{Matt2}.) 
On the other hand, for the  Minkowskian quantities the distribution 
functions $\widehat W_X(\tau)$ are piece-wise analytic, with two 
pieces defined for $|\tau|<1$ and $|\tau|>1$ \cite{BBB}.

\section{Borel sums in the complex momentum plane}
\label{sec:3}

Borel-summed expressions for the polarization function at both 
space-like and time-like momenta in terms of the distribution 
functions $\widehat w_D(\tau)$ and $\widehat W_D(\tau)$ were 
derived in \cite{Matt1} and \cite{BBB}, respectively. In what 
follows, we will generalize the techniques used in these works to 
the calculation of the polarization function in the complex 
momentum plane. For definiteness, we adopt as a regularization 
prescription for the ill-defined integral (\ref{Laplace})
the so-called ``principal value'', which amounts to taking one 
half of the sum of the integrals along two parallel lines 
slightly above and below the real axes, denoted by ${\cal C_\pm}$ 
(the same integration lines appeared previously in (\ref{wn1})). 
We thus define
\begin{equation}\label{PV}
   D(s) = 1 + \frac12\,\Big[ d_+(s) + d_-(s) \Big] \,,
\end{equation}
where
\begin{equation}\label{d+-}
   d_\pm(s) = \frac{1}{\beta_0} \int_{\cal C_\pm}\!{\rm d}u
   \left( \frac{-s}{\Lambda_V^2} \right)^{-u} \widehat B_D(u) \,,
\end{equation}
and $s$ is taken to lie outside the Landau region, $|s|>\Lambda_V^2$, 
so that the integrals are convergent. The prescription (\ref{PV}) is a 
generalization of the Cauchy principal value for simple poles, which 
was adopted in \cite{Khuri1} as the only choice giving a real value 
for the Borel-summed amplitude when the coupling constant is real. In the 
renormalization-group improved expansion, this means that $D(s)$ is real 
along the space-like $s$ axis outside the Landau region. This is 
consistent with the dispersion relation (\ref{dispreld}), which follows 
from general causality and unitarity requirements. As we shall see, 
other prescriptions violate these requirements.

Our aim is to express the integrals (\ref{d+-}) in terms of the 
distribution functions defined in the previous section. To this end,
we must pass from the integrals along the contours ${\cal C_\pm}$ to 
integrals along a line parallel to the imaginary axis, where the 
representation (\ref{wninv}) is valid. This can be achieved 
by rotating the integration contour from the real to the imaginary 
axis, provided the contour at infinity does not contribute. In each 
individual case, one must establish that this condition is satisfied. 
Let us consider first a point in the upper half of the momentum plane,
for which $s=|s|\,e^{i\phi}$ and $-s=|s|\,e^{i(\phi-\pi)}$ with a phase 
$0<\phi<\pi$. Taking $u={\cal R}\,e^{i\theta}$ on a large semi-circle 
of radius ${\cal R}$, the relevant exponential appearing in the 
integrals (\ref{d+-}) is
\begin{equation}\label{expon}
   \exp\left\{ -{\cal R} \left[ \ln\frac{|s|}{\Lambda_V^2}\cos\theta 
   + (\pi-\phi)\sin\theta \right] \right\} \,.
\end{equation}
For $|s|>\Lambda_V^2$, the exponential is negligible at large ${\cal R}$ 
for $\cos\theta>0$ and $\sin\theta>0$, i.e., for the first quadrant of 
the complex $u$ plane. Assuming that $\widehat B_D(u)$ increases 
slower than any exponential \cite{BBB}, the integration contour 
defining $d_+(s)$ can be rotated to the positive imaginary axis, where
the representation (\ref{wninv}) is valid. This leads to the double
integral
\begin{equation}\label{double}
   d_+(s) = \frac{1}{\beta_0} \int\limits_0^{i\infty}\!{\rm d}u
   \int\limits_0^\infty\!{\rm d}\tau\,\widehat w_D(\tau) \exp\left[
   -u\left( \ln\frac{\tau|s|}{\Lambda_V^2} + i(\phi-\pi) \right)
   \right] \,.
\end{equation}
The order of integrations over $\tau$ and $u$ can be interchanged,
since for negative $(\phi-\pi)$ the integral over $u$ is convergent. 
Performing this integral yields
\begin{equation}\label{dplus}
   d_+(s) = \frac{1}{\beta_0} \int\limits_0^\infty\!{\rm d}\tau\,
   \frac{\widehat w_D(\tau)}{\ln(-\tau s/\Lambda_V^2)} \,.
\end{equation}
Consider now the evaluation of the function $d_-(s)$ given by 
the integral along the contour ${\cal C_-}$ below the real axis. 
Naively, one might think to rotate the integration contour to the 
negative imaginary axis without crossing any singularities. However, 
this rotation is not allowed, because along the corresponding 
quarter-circle $\sin\theta<0$, and the exponent (\ref{expon}) does not 
vanish at infinity for $\phi<\pi$. The way out is to perform again a 
rotation to the positive imaginary $u$ axis, for which the 
contribution of the circle at infinity vanishes. But in this rotation 
the contour crosses the positive real axis, and hence we must pick up 
the contributions of the IR renormalon singularities located along 
this line. This is easily achieved by comparing the expression 
(\ref{wn1}) for the function $\widehat w_D^{(<)}(\tau)$ with the 
definition (\ref{d+-}) of the functions $d_\pm(s)$. It follows that 
$d_-(s)$ can be expressed in terms of $d_+(s)$ as
\begin{equation}\label{dif}
   d_-(s) = d_+(s) - \frac{2\pi i}{\beta_0} \left(
   -\frac{\Lambda_V^2}{s} \right)
   \widehat w_D^{(<)}(-\Lambda_V^2/s) \,.
\end{equation}
The relations (\ref{dplus}) and (\ref{dif}) completely specify the
function $D(s)$ in the upper half of the momentum plane. Using the 
same method, the Adler function can be calculated in the lower half 
plane, which corresponds to taking $\pi<\phi<2\pi$. In this case, the 
integral along ${\cal C_-}$ can be calculated by rotating the contour 
up to the negative imaginary $u$ axis, while for the integration 
along ${\cal C_+}$ one must pass across the real axis. Combining the 
results, we obtain the following expression for the Adler function in 
the complex momentum plane:
\begin{equation}\label{duplo}
   D^{(\pm)}(s) = 1 + \frac{1}{\beta_0} \int\limits_0^\infty\!
   {\rm d}\tau\,\frac{\widehat w_D(\tau)}{\ln(-\tau s/\Lambda_V^2)} 
   \mp \frac{i\pi}{\beta_0} \left( -\frac{\Lambda_V^2}{s} \right)
   \widehat w_D^{(<)}(-\Lambda_V^2/s) \,,
\end{equation}
where the superscript ``$\pm$'' in parenthesis refers to the sign of 
$\mbox{Im}\,s$. The corresponding expression for the polarization 
function $\Pi(s)$ can be obtained by inserting the above result into 
the definition (\ref{Pidef}). This gives
\begin{equation}\label{piuplo}
   \Pi^{(\pm)}(s) = k - \ln\left( \frac{-s}{\Lambda_V^2} \right)
   - \frac{1}{\beta_0} \int\limits_0^\infty\!{\rm d}\tau\,
   \widehat w_D(\tau) \ln\ln\left( -\frac{\tau s}{\Lambda_V^2}
   \right) \mp \frac{i\pi}{\beta_0}\left(- \frac{\Lambda_V^2}{s}\right)
\widehat w_{\Pi}^{(<)}(-\Lambda_V^2/s) \,,
\end{equation}
with $\widehat w_\Pi(\tau)$ as defined in (\ref{wdwpi}). Note that the
last terms in (\ref{duplo}) and (\ref{piuplo}) involve the analytic 
continuation of the Mellin transforms $\widehat w_D(\tau)$ and 
$\widehat w_\Pi(\tau)$ from the real positive axis, where they have 
been calculated, to arbitrary complex arguments.

Before discussing in the next section the analyticity properties of 
these results, we apply the same technique to the principal-value 
Borel summation of Minkowskian quantities. Let us generically denote 
a Minkowskian quantity by $R(s_0)$, with $s_0$ some fixed scale, and 
write its Laplace integral with the principal-value prescription as
\begin{equation}\label{PVR}
   R(s_0) = R_0 + \frac12\,\Big[ r_+(s_0) + r_-(s_0) \Big] \,,
\end{equation}
where $R_0$ is a constant, and
\begin{equation}\label{r+-}
   r_\pm(s_0) = \frac{1}{2i\beta_0} \int_{\cal C_\pm}\!{\rm d}u\, 
   \widehat b_R(u) \left\{ \exp\left[ -u\ln\left(
   \frac{s_0}{\Lambda_V^2} \right) + i\pi u \right]
   - \exp\left[ -u\ln\left( \frac{s_0}{\Lambda_V^2} \right)
   - i\pi u \right] \right\} \,.
\end{equation}
We have expressed the Borel transform $\widehat B_R(u)$ in terms of 
$\widehat b_R(u)$ as in (\ref{Bb}) and combined the two exponentials 
arising from the factor $\sin\pi u$ with the exponential in the Laplace 
integral. We recall that $s_0>\Lambda_V^2$. The procedure described in 
detail for complex $s$ can be now applied in a straightforward way. For 
${\cal C_+}$ (${\cal C_-}$) one can rotate the integral of the first 
(second) term in (\ref{r+-}) towards the positive (negative) imaginary 
axis, where the integral representation (\ref{wbbinv}) of 
$\widehat b_R(u)$ in terms of $\widehat W_R(\tau)$ can be used. For the 
remaining terms, i.e., the second (first) term in the integral along 
${\cal C_+}$ (${\cal C_-}$), the simple rotation towards the imaginary 
axis cannot be performed, since the corresponding terms $\pm i\pi u$ in 
the exponential blow up along the quadrants of the circle. As explained 
before, we must first cross the real axis in the $u$ plane and then 
make a rotation towards the nearest imaginary semi-axis. But in 
crossing the real axis we encounter the IR renormalons, whose 
contribution must be added using an analog of (\ref{dif}). In the 
present case, the relation is
\begin{eqnarray}\label{difr}
   &&\int_{\cal C_+}\!{\rm d}u\,\widehat b_R(u) \left\{ \exp\left[
    -u\ln\left( \frac{s_0}{\Lambda_V^2} \right) \pm i\pi u \right]
    \right\} \nonumber\\
   &=& \int_{\cal C_-}\!{\rm d}u\,\widehat b_R(u) \left\{
    \exp\left[ -u\ln\left( \frac{s_0}{\Lambda_V^2} \right) \pm i\pi u
    \right] \right\} + 2\pi i \tau_\pm \widehat W_R^{(<)}(\tau_\pm) \,,
\end{eqnarray}
where $\tau_\pm=-\Lambda_V^2/s_0\pm i\epsilon$. Using this relation, we 
find after a straightforward calculation
\begin{equation}\label{rpv}
   R(s_0) = R_0 + 
   \frac{\pi}{\beta_0} \int\limits_0^\infty\!{\rm d}\tau\, 
   \frac{\widehat W_R(\tau)}{\ln^2(\tau s_0/\Lambda_V^2) + \pi^2}
   + \frac{\pi}{2\beta_0}\,\frac{\Lambda^2_V}{s_0} \Big[
   \widehat W_R^{(<)}(\tau_+) + \widehat W_R^{(<)}(\tau_-) \Big] \,.
\end{equation}
Notice that the last two terms involve the values of the function
$\widehat W_R^{(<)}(\tau)$ for real negative values of the argument, 
where this function has a cut. In the two terms the cut is approached 
from opposite directions, and the sum gives twice the real part of the 
function $\widehat W_R^{(<)}(\tau)$ at $\tau=-\Lambda_V^2/s_0$. It is
convenient to rewrite the above result using an integration by parts,
yielding
\begin{equation}\label{ibp}
   R(s_0) = R_0 + \frac{1}{\beta_0}
   \int\limits_0^\infty\!\mbox{d}\tau\,I_R(\tau)\,
   \arctan\left[ \frac{\pi}{\ln(\tau s_0/\Lambda^2_V)} \right]
   + \frac{\pi}{\beta_0}\,\mbox{Re}\!
   \int\limits_{-\Lambda^2_V/s_0}^{\Lambda^2_V/s_0}\! 
   \mbox{d}\tau\,I_R(\tau-i\epsilon) \,,
\end{equation}
where 
\begin{equation}
   I_R(\tau) = \frac{\mbox{d}}{\mbox{d}\tau}\,\Big[
   \tau\,\widehat W_R(\tau) \Big] \,.
\end{equation}
The integral representations (\ref{rpv}) and (\ref{ibp}) were derived 
previously in \cite{BBB} by means of a different technique of treating 
the integrals (\ref{r+-}). Specifically, in order to avoid the crossing 
of the real axis in the rotation of the contours ${\cal C_\pm}$ towards 
the imaginary axis, these authors use formally the integral 
representation (\ref{wbbinv}) of $\widehat b_R(u)$ outside its range of 
validity. The double integral thus obtained is evaluated by performing 
first a suitable rotation of the integration line in the $\tau$ plane, 
and then a rotation of the integration contour in the $u$ plane. The 
formal application of this procedure leads to an expression identical 
to (\ref{rpv}). 

For our further discussion, it will be useful to have an explicit
expression for the functions $I_{M_k}(\tau)$ entering the integral 
representation (\ref{ibp}) in the particular case of the Borel sum of 
the spectral moments $M_k$. These functions are given by 
\cite{Matt3}\footnote{In this reference, results are given for the 
functions $W_k(\tau)=\pi\tau\,I_{M_k}(\tau)$.}
\begin{equation}\label{Ik}
   I_{M_k}(\tau) = \frac{k+1}{\pi} \int\limits_0^1\!\mbox{d}x\,
   x^{k-1}\,\widehat w_D(\tau/x) = \frac{k+1}{\pi}\,\tau^k
   \int\limits_\tau^\infty\!\mbox{d}z\,z^{-k-1}\,\widehat w_D(z)
   \,, \quad \tau>0 \,.
\end{equation}
We stress that from the derivation of the result (\ref{ibp}) it
follows that for negative values of $\tau$ the functions 
$I_{M_k}(\tau)$ must be obtained from the analytic continuation of the 
expressions derived from for $\tau$ real and positive. Note that for 
$\tau<0$ these functions are not the same as the functions 
\begin{equation}\label{Ikpr}
   I_{M_k}'(\tau) = \frac{k+1}{\pi}\,(-\tau)^k
   \int\limits_{-\tau}^\infty\!\mbox{d}z\,z^{-k-1}\,\widehat w_D(-z)
   \,, \quad \tau<0
\end{equation}
one would obtain using the analytic continuation of the function 
$\widehat w_D(\tau/x)$ under the integral in (\ref{Ik}). This 
observation will become important in the next section.

\section{Analyticity properties of the Borel-summed functions}
\label{sec:4}

Let us now investigate in more detail the momentum-plane 
analyticity properties of the Adler function and of the 
polarization function calculated in the previous section. We start
with the Borel sum of the function $D(s)$. From (\ref{duplo}), it is 
apparent that its $s$ dependence is isolated in two terms having a 
different origin. The first term can be regarded as an average over 
gluon virtualities of the one-loop expressions in (\ref{Drengr}) 
\cite{Matt1}. It is obtained using the integral representation of 
the Borel function in terms of the inverse Mellin transform along 
the imaginary axis in the $u$ plane. The second term is generated by 
the residues of the IR renormalons when crossing the real axis to 
pass from one of the contours ${\cal C_\pm}$ to the other. As we 
will see, the fact that this term contains the piece 
$\widehat w_D^{(<)}(\tau)$ of the distribution function has 
important consequences on the analyticity properties. 

As shown in (\ref{duplo}), we obtained two different expressions 
for the Adler function valid in the upper and lower half 
of the momentum plane. Using general theorems on the analyticity of 
functions represented by integrals \cite{Eden,SaZy} and the fact 
that $\widehat w_D^{(<)}(\tau)$ is holomorphic for complex values of 
$\tau$, it follows that the functions $D^{(\pm)}(s)$ are 
holomorphic for complex values of $s$ outside the real axis, in the 
upper and lower half planes, respectively. We will now show that 
they represent a unique function analytic in the cut momentum plane. 
To this end, we apply the Schwarz reflection principle. The 
functions $D^{(\pm)}(s)$ are related to each other by the reality 
condition $[D^{(+)}(s^*)]^*=D^{(-)}(s)$, since $w_D^{(<)}(\tau)$ is 
a real analytic function, which enters the expressions for 
$D^{(+)}(s)$ and $D^{(-)}(s)$ with imaginary coefficients of 
opposite sign. Assume now that $s$ is approaching the real axis from 
the upper half plane. Using expression (\ref{duplo}) together with 
the relation
\begin{equation}\label{A1}
   \mbox{Im}\left[ \frac{1}{\ln[-(x+i\epsilon)]} \right]
   = \theta(x)\,\frac{\pi}{\ln^2(x)+\pi^2} + \pi\delta(x+1) \,,
\end{equation}
we obtain on the time-like axis
\begin{equation}\label{imagdti}
   \mbox{Im}\,D(s+i\epsilon) = \frac{\pi}{\beta_0} \left[\,
   \int\limits_0^\infty\!{\rm d}\tau\,
   \frac{\widehat w_D(\tau)}{\ln^2(\tau s/\Lambda_V^2) +\pi^2}
   + \frac{\Lambda_V^2}{s}\,\mbox{Re}\,\widehat w_D^{(<)}
   (-\Lambda_V^2/s) \right] \,,\quad s>0 \,,
\end{equation}
while on the space-like axis
\begin{equation}\label{imagdsp}
   \mbox{Im}\,D(s+i\epsilon) = \frac{\pi}{\beta_0} \left(
   -\frac{\Lambda_V^2}{s} \right) \left[ \widehat w_D(-\Lambda_V^2/s)
   - \widehat w_D^{(<)}(-\Lambda_V^2/s) \right] \,,\quad s<0 \,.
\end{equation}
We recall that we have assumed that $s$ lies outside the Landau 
region, i.e., $|s|>\Lambda_V^2$. In this case, the argument of the 
functions $\widehat w_D(\tau)$ and $\widehat w_D^{(<)}(\tau)$ 
appearing in (\ref{imagdsp}) is less than one. But for such values 
the two functions coincide, and the two terms in (\ref{imagdsp}) 
compensate each other. Therefore, the imaginary part of the 
Borel-summed Adler function $D(s)$ vanishes for space-like momenta 
outside the Landau region:
\begin{equation}\label{imagdsp1}
   \mbox{Im}\,D(s+i\epsilon) = 0 \,,\quad s<-\Lambda_V^2 \,.
\end{equation}
The same result is obtained when approaching the real negative axis 
from below. Therefore, in the deep Euclidian region $-s>\Lambda_V^2$ 
both expressions in (\ref{duplo}) lead to a real value given by
\begin{equation}\label{deucl1}
   D(s) = 1 + \frac{1}{\beta_0}\,\mbox{Re} \int\limits_0^\infty\!
   {\rm d}\tau\,\frac{\widehat w_D(\tau)}{\ln(-\tau s/\Lambda_V^2)}
   \,,\quad s<-\Lambda_V^2 \,.
\end{equation}
By the Schwarz reflexion principle, the two expressions $D^{(\pm)}(s)$ 
in (\ref{duplo}) define a single function $D(s)$ real analytic in the 
momentum plane, with no cut for $s<-\Lambda_V^2$. 

Before proceeding, it is worth emphasizing that the vanishing of the 
imaginary part on the space-like $s$ axis outside the Landau region 
is not a generic feature of the Borel sum for the Adler function, but 
is specific to the principal-value prescription adopted above. It does 
not happen if one takes a different prescription for avoiding the 
singularities of the Laplace integral. For instance, taking the 
integral along the contour ${\cal C_+}$ as a definition of the Borel 
integral we would obtain instead of (\ref{duplo}) the expressions
\begin{eqnarray}\label{duplo1}
   D^{(+)}_+(s) &=& 1 + \frac{1}{\beta_0} \int\limits_0^\infty\!
    {\rm d}\tau\,\frac{\widehat w_D(\tau)}{\ln(-\tau s/\Lambda_V^2)}
    \,, \nonumber\\
   D^{(-)}_+(s) &=& 1 + \frac{1}{\beta_0} \int\limits_0^\infty\!
    {\rm d}\tau\,\frac{\widehat w_D(\tau)}{\ln(-\tau s/\Lambda_V^2)}
    + \frac{2\pi i}{\beta_0}\left( -\frac{\Lambda_V^2}{s} \right)
    \widehat w_D^{(<)}(-\Lambda_V^2/s) \,,
\end{eqnarray}
where the subscript ``+'' indicates the regularization prescription. 
It is obvious that these expressions do not satisfy the condition of
real analyticity $D^*(s)=D(s^*)$. In particular, using the relation 
(\ref{A1}) we see  that the two expressions in (\ref{duplo1}) have 
non-zero imaginary parts along the whole space-like axis $s<0$. 
(These imaginary parts coincide, while the real parts exhibit a 
discontinuity across the real axis.) On the other hand, for real 
analytic functions the discontinuity is due to the imaginary part, 
since in this case $D(s+i\epsilon)-D(s-i\epsilon)
=2i\,\mbox{Im}\,D(s+i\epsilon)$. By unitarity, the discontinuity of 
the exact function $D(s)$ is non-zero only above the threshold for 
hadron production, and below this threshold the exact function must be 
real.\footnote{These properties, which follow from the general
principles of field theory, are explicitely incorporated in the 
dispersion relation (\protect\ref{dispreld}), due to the fact that the 
spectral function ${\rm Im}\,\Pi(s)$ is real.} 
The expressions in (\ref{duplo1}) are in conflict with the above 
requirements, and in our opinion this fact is an argument in favour of 
the principal-value prescription.

The explicit expression for $D(s)$ derived in (\ref{duplo}) for 
$|s|>\Lambda_V^2$ can be analytically continued inside the circle 
$|s|<\Lambda_V^2$. It is interesting to make this continuation in 
order to explore what happens, after Borel summation, with the Landau 
pole present in the fixed-order perturbative expression
(\ref{Drengr}). We find that for complex values of $s$ outside the real 
axis the resulting expression is holomorphic. The discontinuities along 
the real axis can be analysed using (\ref{imagdti}) and (\ref{imagdsp}). 
On the time-like axis, we find a rather complicated behaviour
besides the expected unitarity branch point at $s=0$ produced by the 
logarithm. Indeed, the spectral function at $s<\Lambda_V^2$ involves the 
values of the function $\widehat w_D^{(<)}(\tau)$ extended analytically 
in the range $|\tau|>1$, where there may be additional singularities 
(see, e.g., the explicit expressions (\ref{wDfun}) valid in the 
large-$\beta_0$ limit). Singularities are also present along the 
space-like region $-\Lambda_V^2<s<0$. Indeed, in this case the argument 
of the functions $\widehat w_D(\tau)$ and $\widehat w_D^{(<)}(\tau)$ in 
(\ref{imagdsp}) is $\Lambda_V^2/|s|>1$. Therefore, the last two terms 
in (\ref{imagdsp}) do not compensate each other anymore, and hence the 
analytic continuation of the expression for $D(s)$ derived above has a 
cut along the low-energy Euclidian region with the discontinuity
\begin{equation}\label{imagdsp2}
   \mbox{Im}\,D(s+i\epsilon) = \frac{\pi}{\beta_0} \left(
   -\frac{\Lambda_V^2}{s} \right) \left[ \widehat w_D^{(>)}
   (-\Lambda_V^2/s) - \widehat w_D^{(<)}(-\Lambda_V^2/s) \right] \,,
   \quad -\Lambda_V^2 < s < 0 \,.
\end{equation}
The appearance of this cut is explicitely related to the fact that the 
inverse Mellin transform of the Borel function is a piece-wise analytic 
function. Note that even if IR renormalons were absent and the theory 
was Borel summable, the Landau cut would persist. Indeed, in this case  
$d_+(s)=d_-(s)$, and from (\ref{dif}) it follows that the function 
$\widehat w_D^{(<)}(\tau)$ vanishes identically. But in (\ref{imagdsp2}) 
we are then left with the term $\widehat w_D^{(>)}(\tau)$, which 
receives contributions from the UV renormalons. 

The presence of the Landau cut (\ref{imagdsp2}) is intimately related 
with the asymptotic behaviour of the Borel transform. It is 
straightforward to show that if the Borel transform vanishes 
sufficiently fast at infinity in the complex $u$ plane -- more 
precisely, if not only $\widehat B_D(u)$ but also the product 
$\widehat B_D(u)\,\sin\pi u$ are $L^2$ integrable functions in the sense 
of (\ref{L2B}) -- then the Landau cut is absent. Indeed, in this case 
a reasoning similar to that presented in the discussion containing 
equations (\ref{Bb})--(\ref{rel}) leads to the conclusion that the 
inverse Mellin transform $\widehat w_D(\tau)$ satisfies a dispersion 
representation and is therefore analytic in the complex $\tau$ plane cut 
along the negative axis. Then the two pieces $\widehat w_D^{(<)}(\tau)$ 
and $\widehat w_D^{(>)}(\tau)$ coincide, and the imaginary part 
(\ref{imagdsp2}) vanishes identically. However, the strong decrease of 
the Borel transform ensuring the absence of the Landau cut is not 
observed in the large-$\beta_0$ limit, and it is thus rather 
improbable that it should happen in the physical case.

As a side remark, let us indicate what happens if the effective 
coupling constant is modified in the IR domain so as not to contain a 
Landau pole. To study this case we make the replacement
\begin{equation}\label{repl}
   -\frac{\Lambda_V^2}{s} ~\to~ \tau_L \equiv
   \exp\left( -\frac{4\pi}{\beta_0\,\alpha_s(-s)} \right) \,,
\end{equation}
where the coupling $\alpha_s(-s)$ is defined in the V scheme. Then the 
discontinuity (\ref{imagdsp}) across the Landau cut may be written as
\begin{equation}\label{imagdspan}
   \mbox{Im}\,D(s+i\epsilon) = \frac{\pi}{\beta_0}\,\tau_L \left[
   \widehat w_D(\tau_L) - \widehat w_D^{(<)}(\tau_L) \right]
   \,,\quad s<0 \,.
\end{equation}
The coupling $\alpha_s(-s)$ may be defined 
by a dispersion integral \cite{DoMa} or simply by subtracting the 
pole at $s=-\Lambda_V^2$ by hand. As discussed in \cite{BBB}, this 
redefinition amounts to adding terms exponentially small in the coupling, 
which do not modify the perturbative expansion. The modified coupling is 
real and positive along the whole space-like region $s<0$, implying that 
$\tau_L$ defined in (\ref{repl}) is a positive number less than unity. 
As a consequence, the imaginary part of $D(s)$ vanishes along the  
space-like axis, since for $\tau_L <1$ the two terms in (\ref{imagdspan}) 
compensate each other. Therefore, unphysical singularities do not appear 
in the Borel-summed expansion in powers of a regular coupling, if the 
principal-value prescription is adopted for treating the IR 
renormalons.\footnote{This argument cannot be considered a proof, 
since the very meaning of the Borel summation of a 
modified expansion with the Landau pole removed is not transparent. 
Indeed, the redefinition of the coupling itself may be seen as a 
partial summation of some higher-order effects, which must be 
separated from the terms taken into account in the Borel sum in order 
to avoid double counting.}

For completeness, we also present results for the polarization 
function $\Pi(s)$. The Borel-summed expression   
(\ref{piuplo}), obtained for $|s|>\Lambda_V^2$, can be 
analytically continued to the whole complex plane. We note that 
$\Pi(s)$ is holomorphic for complex values of $s$ and satisfies the 
reality condition $\Pi(s^*)=\Pi^*(s)$. On the real axis, this function 
can have singularities manifested as discontinuities of the imaginary 
part. A straightforward calculation gives
\begin{eqnarray}\label{imagpi1}
   \mbox{Im}\,\Pi(s+i\epsilon) &=& \pi + \frac{1}{\beta_0}
    \int\limits_0^\infty\!{\rm d}\tau\,\widehat w_D(\tau)
    \arctan\left(\frac{\pi}{\ln(\tau s/\Lambda_V^2)} \right)
    \nonumber\\
   &&\mbox{}+ \frac{\pi}{\beta_0}\,\mbox{Re}\!
    \int\limits_{-\Lambda_V^2/s}^{\Lambda_V^2/s}\!{\rm d}\tau\,
    \widehat w_D(\tau) \,,\quad s>0 \,, \nonumber\\
   \mbox{Im}\,\Pi(s+i\epsilon) &=& \frac{\pi}{\beta_0} \left(
    - \frac{\Lambda_V^2}{s} \right) \left[ \widehat w_\Pi(-\Lambda_V^2/s)
    - \widehat w_\Pi^{(<)}(-\Lambda_V^2/s) \right] \,,\quad s<0 \,.
\end{eqnarray}
Again, the second expression vanishes outside the Landau region, where 
the function $\widehat w_\Pi(\tau)$ coincides with 
$\widehat w_\Pi^{(<)}(\tau)$. However, inside the interval 
$-\Lambda_V^2<s<0$ the function $\Pi(s)$ has a non-zero imaginary 
part.

The final point we will discuss concerns the implications of the above 
results for the Borel summation of Minkowskian quantities such as the 
spectral moments $M_k$. In (\ref{ibp}) we derived the principal-value
Borel sum for these quantities, which we shall denote by  
$M_k^{\rm Borel}$. On the other hand, starting from the explicit 
expression (\ref{piuplo}) for the polarization function $\Pi(s)$ in 
the complex plane, we could calculate its moments also by performing the 
contour integral in the second relation in (\ref{Mkdef}). Let us denote 
by $M_k^{\rm circle}$ the result of this procedure. We are now going to
show that $M_k^{\rm Borel}=M_k^{\rm circle}$, which is equivalent to 
the statement that contour integration and Borel summation commute 
with each other. In order to compare the two resummed 
expressions for the moments, we pass from the integral along the 
circle $|s|=s_0$ to an integral along the real axis. Applying the 
Cauchy relation, and taking into account the singularities of the 
Borel sum of $\Pi(s)$ inside the circle, we write in the most general 
way
\begin{eqnarray}\label{Cauchy}
   M_k^{\rm circle} &\stackrel{\eta\to 0}{=}&
    - \frac{k+1}{2\pi i\,s_0^{k+1}} \oint\limits_{|s|=\eta}\!
    \mbox{d}s\,s^k\,\Pi(s) \nonumber\\
   &&\mbox{}+ \frac{k+1}{\pi\,s_0^{k+1}} \left[
    \int\limits_{-s_0}^{-\eta}\!\mbox{d}s\,s^k\,\mbox{Im}\,
    \Pi(s+i\epsilon) + \int\limits_\eta^{s_0}\!
    \mbox{d}s\,s^k\,\mbox{Im}\,\Pi(s+i\epsilon) \right] \,.
\end{eqnarray}
The point $s=0$ must be treated separately since the function 
$\widehat w_\Pi^{(<)}(-\Lambda^2_V/s)$ entering the expression
for $\Pi(s)$ in (\ref{piuplo}) is singular at this point. (In the 
large-$\beta_0$ limit, we have $\widehat w_\Pi^{(<)}(\tau)\sim\tau 
\ln^2\!\tau$ for $\tau\to\infty$.) From this behaviour, it follows 
that the circle of radius $\eta$ gives a non-vanishing contribution 
only for the first two moments. For $k\ge 2$, the only singularities 
are due to the discontinuity of the imaginary part of $\Pi(s)$ given 
above. As we discussed, this imaginary part vanishes for 
$s<-\Lambda_V^2$. By a straightforward calculation, using 
(\ref{imagpi1}) and isolating  the contribution of the region 
$-\Lambda_V^2<s<\Lambda_V^2$, we write (\ref{Cauchy}) as
$M_k^{\rm circle}=M_k'+\delta_k$, where
\begin{eqnarray}\label{deltamk}
   \delta_k &\stackrel{\eta\to 0}{=}& -\frac{k+1}{2\pi i\,s_0^{k+1}}
    \oint\limits_{|s|=\eta}\!\mbox{d}s\,s^k\,\Pi(s) \nonumber\\
   &&\mbox{}+ \frac{k+1}{\beta_0\,s_0^{k+1}} \left[
    \int\limits_{-\Lambda^2_V}^{-\eta}\!\mbox{d}s\,s^k\,
    \Delta(-\Lambda^2_V/s)
    + \int\limits_\eta^{\Lambda^2_V}\!\mbox{d}s\,s^k\,
    \mbox{Re}\,\Delta(-\Lambda^2_V/s) \right] \,,
\end{eqnarray}
with
\begin{equation}\label{delta}
   \Delta(\tau)= \tau \left[ \widehat w_\Pi^{(>)}(\tau)
   - \widehat w_\Pi^{(<)}(\tau) \right] \,.
\end{equation}
The quantity $M_k'$ has an expression that is formally identical with 
the Borel moments $M_k^{\rm Borel}$ in (\ref{ibp}), but in which the 
functions $I_{M_k}(\tau)$ in (\ref{Ik}) are replaced with the 
functions $I_{M_k}'(\tau)$ in (\ref{Ikpr}) obtained using a different 
prescription for negative $\tau$. In other words, we have
$M_k'=M_k^{\rm Borel}-\delta_k'$ with
\begin{equation}\label{difik}
   \delta_k' = \frac{\pi}{\beta_0}\,\mbox{Re}\!
   \int\limits_{-\Lambda_V^2/s_0}^0\!\mbox{d}\tau\,
   \Big[ I_{M_k}(\tau-i\epsilon) - I_{M_k}'(\tau-i\epsilon) \Big] \,.
\end{equation}
An explicit calculation in the large-$\beta_0$ limit shows that 
$\delta_k=\delta_k'$, and we believe this result is of general 
validity. From this observation, it follows that indeed 
$M_k^{\rm circle}=M_k^{\rm Borel}$, which is in agreement with the
findings of \cite{BBB}. Note that the arguments given above show that 
the equivalence of the two resummation procedures results from a 
subtle compensation of the contribution of the singularities of 
$\Pi(s)$ along the Landau region with an additional term arising from 
the analytic continuation of a piece-wise analytic function.

\section{Conclusions}
\label{sec:5}

We have derived a compact expression for the Borel-summed Adler 
function and the polarization amplitude in the complex momentum 
plane in terms of the inverse Mellin transform of the corresponding 
Borel functions. At present, these inverse Mellin transforms 
(or distribution functions) are known only in the large-$\beta_0$ 
limit. However, the expressions derived in the present work are 
useful for investigating the analyticity properties in the complex 
momentum plane, and therefore some conclusions can be drawn even if 
the distribution functions are not known exactly.
 
Our main results are the expressions for the Borel sums of the 
functions $D(s)$ and $\Pi(s)$ given in (\ref{duplo}) and (\ref{piuplo}), 
respectively. They have been derived using a generalized 
principal-value prescription, which leads to real values of the 
functions along the space-like $s$ axis outside the Landau region, 
in accordance with general analyticity requirements imposed by 
causality and unitarity. The expressions derived in the momentum 
plane outside the Landau region admit an explicit analytical 
continuation to the region $|s|<\Lambda_V^2$, revealing Landau 
singularities more complicated than those present in fixed-order 
perturbation theory. We have shown that the discontinuity across the 
Landau cut is connected with the piece-wise character of the inverse 
Mellin transform, which is in turn correlated with the asymptotic 
behaviour of the Borel transform. We have also presented a condition 
on the asymptotic behaviour of the Borel transform that would ensure 
the vanishing of the Landau cut. This condition is, however, not 
satisfied in the large-$\beta_0$ limit, and it will most likely not 
be satisfied for real QCD. Therefore, we do expect the Landau cut to
be a general feature of Borel-summed perturbation theory in QCD.
 
With the explicit expressions derived in the complex momentum plane, 
we have checked that the spectral moments defined by contour 
integrals of the Borel sum of the correlator $\Pi(s)$ coincide with 
the ``standard'' Borel moments considered previously in 
\cite{Max}--\cite{Matt3}. This equality is consistent with the 
complicated singularity structure of the polarization function 
discovered in this work, providing an indirect check of our results. 

We have shown that the unphysical discontinuity across the Landau cut 
would persist in the fictitious case of a Borel-summable perturbation 
series. Thus, the Landau singularities in the momentum plane and the 
IR renormalons are not directly related to each other. A similar 
conclusion was formulated in the context of a specific model in 
\cite{DoUr}. The unphysical cut is a manifestation of the 
incompleteness of any perturbative approximation to a physical 
hadronic quantity. On the other hand, our results suggest that, if 
the running coupling is regularized by a dispersion relation, then
the analyticity properties of fixed-order perturbation theory are 
preserved by Borel summation. Indeed, in this case the discontinuity 
across the Landau cut vanishes even if IR renormalons are present. 
This result holds under conservative assumptions about the properties 
of the inverse Mellin transform and thus seems to have a general 
validity.

\newpage
\acknowledgments
We are grateful to M.~Beneke, C.~Bourrely, P.~Dita, J.~Fischer and 
L.~Lellouch for many interesting discussions. One of us (I.C.) is 
pleased to thank Prof.\ S.~Randjbar-Daemi for his kind hospitality at 
the High Energy Section of ICTP Trieste, where part of this work was 
done. The research of I.C.\ was partially supported by the Ministry 
of Research and Technology, Bucharest, under the Grant No.~3036GR/1997.

\section*{Appendix}
\appendix

Here we indicate a simple method for calculating the distribution 
functions of various observables using the relations between their 
Borel transforms. To this end, we use Parseval's theorem for the Mellin 
transform \cite{Tit}, which can be stated as follows: if ${\cal F}(u)$ 
and ${\cal G}(u)$ satisfy the $L^2$ conditions
\begin{equation}\label{L2F}
   \frac{1}{2\pi i} \int\limits_{u_0-i\infty}^{u_0+i\infty}\!
   {\rm d}u\,|{\cal F}(u)|^2 < \infty \,, \qquad
   \frac{1}{2\pi i} \int\limits_{1-u_0-i\infty}^{1-u_0+i\infty}\!
   {\rm d}u\,|{\cal G}(u)|^2 < \infty \,,
\end{equation}
where $0<u_0<1$, then 
\begin{equation}\label{Pars}
   \frac{1}{2\pi i} \int\limits_{u_0-i\infty}^{u_0+i\infty}\!
   {\rm d}u\,{\cal F}(u)\,{\cal G}(1-u)
   = \int\limits_0^\infty\!{\rm d}x\,f(x)\,g(x) \,,
\end{equation}
where $f(x)$ and $g(x)$ are the inverse Mellin transforms
\begin{equation}\label{fg}
   f(x) = \frac{1}{2\pi i} \int\limits_{u_0-i\infty}^{u_0+i\infty}\!
   {\rm d}u\,{\cal F}(u)\,x^{u-1} \,, \qquad
   g(x) = \frac{1}{2\pi i} \int\limits_{1-u_0-i\infty}^{1-u_0+\infty}
   \!{\rm d}u\,{\cal G}(u)\,x^{u-1} \,.
\end{equation}

As a first application of this result, we use it to prove relation 
(\ref{rel}). We start with relation (\ref{wbb}) and notice that the 
function $\widehat W_X(\tau)$ can be written as
\begin{equation}\label{wbb1}
   \widehat W_X(\tau) = \frac{1}{2\pi i}
   \int\limits_{u_0-i\infty}^{u_0+i\infty}\!{\rm d}u\,
   {\cal F}(u)\,{\cal G}(1-u) \,,
\end{equation}
where
\begin{equation}\label{def}
   {\cal F}(u) = \widehat B_X(u) \,, \qquad
   {\cal G}(u) = \frac{\tau^{-u}}{\sin\pi(1-u)} \,.
\end{equation}
It then follows that $f(x)=\widehat w_X(x)$, and inserting the 
definition (\ref{def}) of the function ${\cal G}(u)$ into (\ref{fg}) 
and making the change of variables $s=1-t$ we find
\begin{equation}\label{gg}
   g(x) = \frac{1}{2\pi i} \int\limits_{u_0-i\infty}^{u_0+\infty}\!
   {\rm d}s\,\frac{\tau^{s-1} x^{-s}}{\sin\pi s} \,.
\end{equation}
This integral can be calculated closing the contour in the complex 
$s$ plane and applying the theorem of residues. For 
$\mbox{Re}\,(\tau/s)>0$ we close the contour in the left half plane 
$\mbox{Re}\,s<0$, while for $\mbox{Re}\,(\tau/s)<0$ we close it in 
the right half plane $\mbox{Re}\,s>0$. In each case, we pick up the 
contributions of the relevant poles situated at $s=\pm n$ with the 
residues
\begin{equation}
   \frac{(-1)^n}{\pi\tau} \left( \frac{\tau}{x} \right)^n \,.
\end{equation}
The contributions of the semi-circles at infinity vanish due to the 
exponentials in the sine function, even for complex values of 
$\tau/x=|\tau/x|\exp(\pm i\psi)$ with $|\psi|<\pi$. Therefore, we 
obtain
\begin{equation}\label{gfin}
   g(x) = \frac{1}{\pi\tau} \sum_{n=1}^\infty (-1)^n \left(
   \frac{\tau}{x} \right)^n = \frac{1}{\pi(x+\tau)} \,.
\end{equation}
Inserting the expressions for the 
functions $f(x)$ and $g(x)$ into the right-hand side of (\ref{Pars}), 
and using the definition of $\widehat W_X(\tau)$, we obtain
\begin{equation}\label{rel1ap}
   \widehat W_X(\tau) = \frac{1}{\pi} \int\limits_0^\infty\!{\rm d}x\,
   \frac{\widehat w_X(x)}{x+\tau} \,,
\end{equation}
where $\tau$ can take arbitrary values, except for real negatives.

As a second example, we derive relation (\ref{wdwpi}), starting from
\begin{equation}\label{wPin}
   \widehat w_\Pi(\tau) = \frac{1}{2\pi i}
   \int\limits_{u_0-i\infty}^{u_0+i\infty}\!{\rm d}u\,
   \frac{\widehat B_D(u)}{u}\,\tau^{u-1} \,,
\end{equation}
where we have used (\ref{BDBPi}). The above result can be rewritten 
as in (\ref{wbb1}), with the identification
\begin{equation}\label{def1}
   {\cal F}(u) = \widehat B_D(u) \,, \qquad
   {\cal G}(u) = \frac{\tau^{-u}}{(1-u)} \,.
\end{equation}
A straightforward calculation gives the inverse Mellin transforms
\begin{equation}\label{ffgg}
   f(x) = \widehat w_D(x) \,, \qquad
   g(x) =  \frac{1}{\tau}\,\theta\left( 1 - \frac{x}{\tau} \right) \,.
\end{equation}
Inserting these functions into the right-hand side of Parseval's 
theorem (\ref{Pars}), we obtain relation (\ref{wdwpi}).

\end{document}